\documentclass[%
preprint,
showpacs,preprintnumbers,
 amsmath,amssymb,
 aps,
]{revtex4-1}

\usepackage{graphicx}
\usepackage{dcolumn}
\usepackage{bm}
\usepackage{subfigure} 
\usepackage{color,soul}
\usepackage{ulem}

\begin{document}


\title{Line tension and morphology of a sessile droplet on a spherical substrate}

\author{Masao Iwamatsu}
\email{iwamatsu@ph.ns.tcu.ac.jp}
\affiliation{%
Department of Physics, Tokyo City University, Setagaya-ku, Tokyo 158-8557, Japan\footnote{Permanent address}
}%
\affiliation{
Department of Physics, Tokyo Metropolitan University, Hachioji, Tokyo 192-0397, Japan
}%



\date{\today}

\begin{abstract}
The effects of line tension on the morphology of a sessile droplet placed on top of a convex spherical substrate are studied.  The morphology of the droplet is determined from the global minimum of the Helmholtz free energy. The contact angle between the droplet and the spherical substrate is expressed by the generalized Young's formula.  When the line tension is positive and large, the contact angle jumps discontinuously to $180^{\circ}$, the circular contact line shrinks towards the top of the substrate, and the droplet detaches from the substrate, forming a spherical droplet if the substrate is hydrophobic (i.e., the Young's contact angle is large).  This finding is consistent with that predicted by Widom [J. Phys. Chem. {\bf 99}, 2803 (1995)]; the line tension induces a drying transition on a flat substrate. On the other hand, the contact angle jumps to $0^{\circ}$, the circular contact line shrinks towards the bottom of the substrate, and the droplet spreads over the substrate to form a wrapped spherical droplet if the substrate is hydrophilic (i.e., the Young's contact angle is small). Therefore, not only the drying transition of a cap-shaped to a detached spherical droplet but also the wetting transition of a cap-shaped to a wrapped spherical droplet could occur on a spherical substrate as the surface area of the substrate is finite.  When the line tension is negative and its magnitude increases, the contact line asymptotically approaches the equator from either above or below.  The droplet with a contact line that coincides with the equator is an isolated, singular solution of the first variational problem.  In this instance, the contact line is pinned and cannot move as far as the line tension is smaller than the critical magnitude, where the wetting transition occurs. 

\end{abstract}

\maketitle

\section{\label{sec:sce1}Introduction}
Line tension~\cite{deGennes1985,Dietrich1988,Bonn2009,Bormashenko2013} occurs in the presence of a three-phase contact line, which separates three phases of matter, such as the liquid, solid, and vapor, of a pure substance.  Therefore, despite the fact that researchers have debated whether line tension plays a role in wetting as its overall magnitude is quite low~\cite{Pompe2000,Wang2001,Checco2003,Schimmele2007,Bonn2009}, line tension must play some role for a small cap-shaped droplets wetting a substrate.  In fact, the line tension is known to play a fundamental role in the stability of such a droplet~\cite{Navascues1981,Widom1995,Singha2015,Iwamatsu2015}.

Two decades ago, Widom~\cite{Widom1995} predicted that line tension would induce a morphological transition of a droplet placed on a flat substrate.  There have also been several studies~\cite{Lipowsky2001,Blecua2006} of such line-tension-induced morphological transitions.  Recently, the interest of researchers has turned from flat, structureless substrates to more complex substrates with differing complexities and geometries.  In particular, the wetting and spreading strategies borrowed from biological structures have potential in the development and design of new materials based on the design principle known as biomimetics~\cite{Nosonovsky2007,Song2014}.  However, to date, the line-tension effects have been primarily considered on flat substrates~\cite{Navascues1981,Widom1995}.  There have been almost no theoretical attempts to clarify the line-tension effects on various substrates with complex geometries except for a very small number of works expounding on line-tension effects on spherical substrates~\cite{Guzzardi2007,Hienola2007,Cooper2007,Iwamatsu2015}.  There have also only been a small number of experimental studies concerning  a droplet on a spherical substrate~\cite{Tao2011,Extrand2012}.

In the present study, we extend our previous work~\cite{Iwamatsu2015,Iwamatsu2015b} on the line-tension effects on convex and concave spherical substrates and consider the line-tension effects on the morphology of a cap-shaped droplet of a non-volatile liquid placed on top of a complete sphere.  We find that the equator of the spherical substrate plays a special role.  The contact line of the droplet cannot cross the equator while changing the magnitude of line tension continuously.  When the substrate is hygrophobic or hydrophobic (i.e. Young's contact angle is large), a morphological transition from a cap-shaped droplet to a spherical droplet is observed for a large, positive line tension; this result is similar to the drying transition induced by positive line tension on a flat substrate, as predicted by Widom~\cite{Widom1995}.  When the substrate is hygrophilic or hydrophilic (i.e., the Young's contact angle is small), a wetting transition is observed, in which the droplet completely encloses the spherical substrate to form a spherically wrapped droplet.  The droplet also cannot spread over the substrate indefinitely given a large, negative line tension.  Instead, the contact line asymptotically approaches the equator from either above or from below while increasing the magnitude of negative line tension.  Therefore, the equator plays the same role as infinity for a plane substrate.

In the following, we will study the effect of the line tension on the morphology of a thermodynamically stable and metastable droplet placed on top of a convex spherical substrate using the capillary model.  The stability of a cap-shaped droplet against fluctuations that do not preserve its circular shape will not be considered.  It is well known, however, that the capillary model possesses a short-wavelength instability~\cite{Dobbs1999,Brinkmann2005,Guzzardi2006} on a flat substrate, which was shown~\cite{Mechkov2007} to be not physical when the molecular interaction near the three-phase contact line is included through a disjoining pressure~\cite{Indekeu1992}.  We will leave this problem of the fluctuation and the inclusion of the disjoining pressure on a spherical substrate for future investigation.

\section{\label{sec:sec2}Line-tension effects on the Helmholtz free energy}

\begin{figure}[htbp]
\begin{center}
\includegraphics[width=0.50\linewidth]{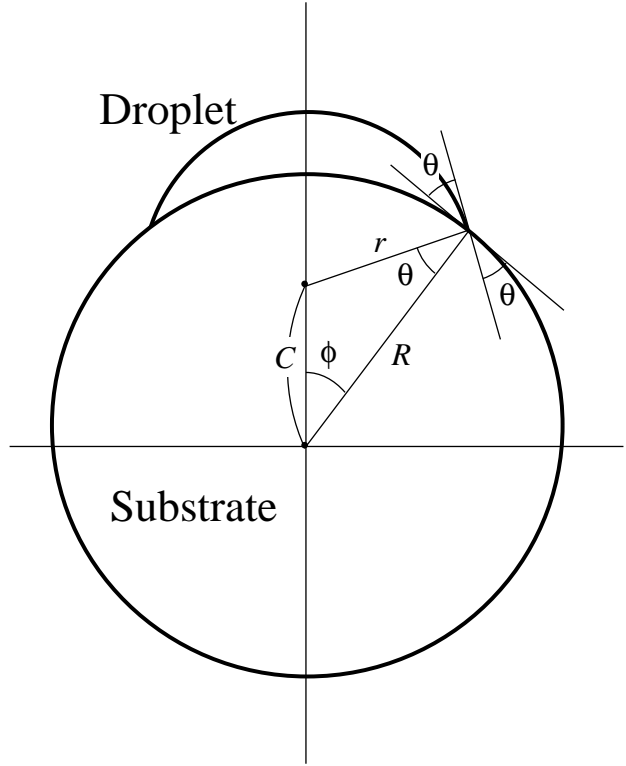}
\caption{
A liquid droplet on a spherical substrate.  The centers of the droplet with radius $r$ and that of the spherical substrate with radius $R$ are separated by a distance $C$.  The contact angle is denoted by $\theta$. Note that the three-phase contact line passes through the equator when $\phi=90^{\circ}$.   }
\label{fig:1L}
\end{center}
\end{figure}

In our previous work~\cite{Iwamatsu2015,Iwamatsu2015b}, we considered line-tension effects on the critical nucleus of a volatile liquid heterogeneously nucleated on a spherical substrate and also considered the Gibbs free energy, which was appropriate to the nucleation.  In this study, we focus on the physics of line tension on a cap-shaped droplet of a non-volatile liquid placed on a spherical substrate, as shown in Fig.~\ref{fig:1L}.  In particular, we consider a droplet with radius $r$ placed on the top of a spherical substrate of radius $R$.  The angle $\theta$ made by two tangential lines at the contact line is the contact angle (Fig.~\ref{fig:1L}).  Since the droplet volume is held constant, the radius $r$ and the contact angle $\theta$ are not independent.  

The number of molecule within the droplet is fixed since we consider an incompressible non-volatile liquid with fixed volume.  Then, we have to consider the Helmholtz free energy of the droplet.  According to the classical idea of wetting and nucleation theory~\cite{Kelton2010,Fletcher1958,Navascues1981,Qian2009}, which is known as the capillary model, the Helmholtz free energy $F$ of a sessile droplet is
\begin{equation}
F=\sigma_{\rm lv}A_{\rm lv}+\Delta\sigma A_{\rm sl}+\tau L,
\label{eq:1L}
\end{equation}
and
\begin{equation}
\Delta\sigma = \sigma_{\rm sl}-\sigma_{\rm sv},
\label{eq:2L}
\end{equation}
where $A_{\rm lv}$ and $A_{\rm sl}$ are the surface areas of the liquid-vapor and liquid-solid (substrate) interfaces, respectively, and $\sigma_{\rm lv}$ and $\sigma_{\rm sl}$ are their respective surface tensions.  Moreover, $\Delta \sigma$ is the free energy gained when the solid-vapor interface with surface tension $\sigma_{\rm sv}$ is replaced by the solid-liquid interface with surface tension $\sigma_{\rm sl}$.   The effect of the line tension $\tau$ is given by the last term of Eq.~(\ref{eq:1L}), where $L$ denotes the length of the three-phase contact line.  When the line tension is positive ($\tau>0$), the droplets tends to shrink to reduce the line length  $L$ and to decrease the free energy $F$.

The contact angle $\theta$ of a mechanically stable droplet is determined by minimizing the Helmholtz free energy
\begin{equation}
F=4\pi R^2\sigma_{\rm lv}f\left(\rho,\theta\right),
\label{eq:3L}
\end{equation}
with
\begin{equation}
f\left(\rho,\theta\right)=\rho\frac{\left(\rho+\zeta\right)^2-1}{4\zeta}-\cos\theta_{\rm Y}\frac{\rho^{2}-\left(1-\zeta\right)^{2}}{4\zeta}+\tilde{\tau}\frac{\rho\sin\theta}{2\zeta},
\label{eq:4L}
\end{equation}
derived from Eq.~(\ref{eq:1L}) with respect to the radius $r$ of the droplet under a condition of constant volume given by
\begin{equation}
V=\frac{4\pi}{3}R^{3}\omega\left(\rho,\theta\right)
\label{eq:5L}
\end{equation}
with
\begin{eqnarray}
\omega\left(\rho,\theta\right)&=&\frac{1}{16\zeta}\left(\zeta-1+\rho\right)^{2} \nonumber \\
&&\times\left[3\left(1+\rho\right)^{2}-2\zeta\left(1-\rho\right)-\zeta^{2}\right]
\label{eq:6L}
\end{eqnarray}
where
\begin{equation}
\zeta=\sqrt{1+\rho^{2}-2\rho\cos\theta}
\label{eq:7L}
\end{equation}
and
\begin{equation}
\rho=\frac{r}{R}
\label{eq:8L}
\end{equation}
is the size parameter of the droplet.  The Young's contact angle $\theta_{\rm Y}$ in Eq.~(\ref{eq:4L}) is defined by the classical Young's equation~\cite{Young1805} on a flat substrate, 
\begin{equation}
\Delta\sigma+\sigma_{\rm lv}\cos\theta_{\rm Y}=0,
\label{eq:9L}
\end{equation}
and the scaled line tension $\tilde{\tau}$ is defined by
\begin{equation}
\tilde{\tau}=\frac{\tau}{\sigma_{\rm lv}R}.
\label{eq:10L}
\end{equation}
Eqs.~(\ref{eq:3L}) and (\ref{eq:6L}) were derived using the integration scheme originally developed by Hamaker~\cite{Hamaker1937}.  The detailed derivation of the volume as well as the Helmholtz energy was given in my previous paper~\cite{Iwamatsu2015} and is also provided in the Appendix.

Extremization~\cite{Navascues1981} of the Helmholtz free energy Eq.~(\ref{eq:3L}) under the
subsidiary constraint of a constant volume, Eq.~ (\ref{eq:5L}), leads to a
relation between the equilibrium contact angle $\theta_{\rm e}$, Young's
angle $\theta_{\rm Y}$, the scaled droplet radius $\rho_{\rm e}$, and scaled line
tension $\tilde{\tau}$, written as
\begin{equation}
\cos\theta_{\rm Y}-\cos\theta_{\rm e}-\tilde{\tau}\frac{1-\rho_{\rm e}\cos\theta_{\rm e}}{\rho_{\rm e}\sin\theta_{\rm e}}=0,
\label{eq:11L}
\end{equation}
which is similar to the classical Young's equation~\cite{Young1805} in Eq.~(\ref{eq:9L}).  Therefore, even on a spherical curved surface, the contact angle in mechanical equilibrium is determined from the classical Young's equation (\ref{eq:9L})~\cite{Fletcher1958,Qian2009} {and $\theta_{\rm e}=\theta_{\rm Y}$ } if line tension can be neglected ($\tau=0$).
Equation~(\ref{eq:11L}) can also be written as
\begin{equation}
\cos\theta_{\rm Y}-\cos\theta_{\rm e}-\frac{\tilde{\tau}}{\tan\phi_{\rm e}}=0
\label{eq:12L}
\end{equation}
where $\phi_{\rm e}$ is the half of the central angle defined in Fig.~\ref{fig:1L}.  Equation (\ref{eq:12L}) is known as the generalized Young's equation~\cite{Hienola2007}.

\begin{figure}[htbp]
\begin{center}
\includegraphics[width=0.50\linewidth]{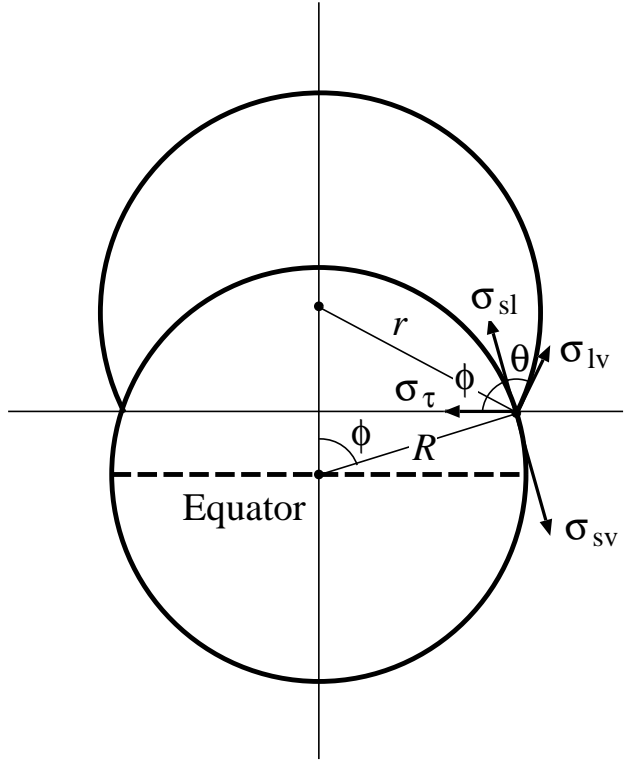}
\caption{
The mechanical-force balance among the three surface tensions $\sigma_{\rm lv}$, $\sigma_{\rm sl}$, and $\sigma_{\rm sv}$ and the tension $\sigma_{\tau}$ from the line tension $\tau$ of the droplet on a convex, spherical substrate. The line-tension contribution $\sigma_{\tau}$ vanishes when $\phi=180^{\circ}$ or $\theta=\theta_{\rm c}$, as given by Eq.~(\ref{eq:15L}). }
\label{fig:2L}
\end{center}
\end{figure}

Equation~(\ref{eq:12L}) can be derived from the mechanical force balance from the surface and line tension, as pointed out by Hienola {\it et al.}~\cite{Hienola2007}.  To this end, we note that the line tension contributes to the force balance (Fig.~\ref{fig:2L}) as~\cite{Iwamatsu2015}
\begin{equation}
\sigma_{\tau}=\frac{\tau}{R\sin\phi}.
\label{eq:13L}
\end{equation}
The line tension contributes to the tension $\sigma_{\tau}$ at the three-phase contact line only when the circular contact line has a finite radius $R\sin\phi$.  Then, a simple force balance between the three tensions $\sigma_{\rm lv}$, $\sigma_{\rm sl}$, and $\sigma_{\rm sv}$ and $\sigma_{\tau}$ (Fig.~\ref{fig:2L}) projected onto the tangential plane leads to
\begin{equation}
\sigma_{\rm sl}-\sigma_{\rm sv}+\sigma_{\rm lv}\cos\theta +\sigma_{\tau}\cos\phi=0,
\label{eq:14L}
\end{equation}
which can be reduced to Eq.~(\ref{eq:12L}). Thus, the contact angle determined from the force balance condition Eq.~(\ref{eq:14L}) is also the equilibrium contact angle $\theta_{\rm e}$ derived from the condition of a free-energy extremum.  Furthermore, a line tension cannot contribute to the balance of tangential stress component at $\phi=\pi/2$, where the contact angle is given by the characteristic contact angle $\theta_{\rm c}$, determined using
\begin{equation}
1-\rho_{\rm c}\cos\theta_{\rm c}=0.
\label{eq:15L}
\end{equation}
derived from Eq.~(\ref{eq:11L}), where $\rho_{\rm c}$ is the radius of the droplet when the contact angle is $\theta_{\rm c}$.  When $\phi=\pi/2$ or $\theta=\theta_{c}$, the three-phase contact line coincides with the equator of the substrate.  In particular, the equilibrium contact angle $\theta_{\rm e}$ in a free energy extremum, as determined from Eq.~(\ref{eq:11L}) is fixed to $\theta_{\rm e}=\theta_{\rm Y}$ when $\theta_{\rm e}=\theta_{\rm c}$ and, therefore, the contact line coincides with the equator, no matter what the magnitude of the line tension $\tilde{\tau}$ is.  The contact angle $\theta_{\rm e}$ of an equilibrium droplet whose contact line coincides with the equator is not affected by the line tension, and is simply given by the Young's contact angle $\theta_{\rm Y}$.  We can know the magnitude of the Young's contact angle $\theta_{\rm Y}$, which characterizes the difference surface energies, by measuring the contact angle of a droplet whose contact line coincides with the equator.

Because the size parameter $\rho=\rho\left(\theta\right)$ is a function of $\theta$ from Eq.~(\ref{eq:6L}) for a fixed volume $V$ or scaled volume $\omega\left(\rho,\theta\right)$, the generalized Young's equation [Eqs.~(\ref{eq:11L}) or (\ref{eq:12L})] determines the contact angle $\theta$. In order to fix the droplet volume, it is convenient to characterize the droplet volume when it is a sphere with the size parameter $\rho_{180}=\rho\left(\theta=180^{\circ}\right)$. Then, we have
\begin{equation}
\omega\left(\rho,\theta\right)=\omega\left(\rho=\rho_{180},\theta=180^{\circ}\right)=\rho_{180}^{3},
\label{eq:16L}
\end{equation}
and, the equilibrium Helmholtz free energy of the droplet with a fixed volume $V$ is obtained as the function of the contact angle $\theta$ from Eq.~(\ref{eq:5L}).  The equilibrium Helmholtz free energy of the cap-shaped droplet at the free-energy extremum (Appendix) is given by
\begin{equation}
F_{\rm cap}=4\pi R^{2}\sigma_{\rm lv}f_{\rm e}
\label{eq:17L}
\end{equation}
with
\begin{eqnarray}
f_{\rm e} &=& \frac{\left(-1+\rho_{\rm e}+\zeta_{\rm e}\right)^{2}\left(\cos\theta_{\rm e}+1+\zeta_{\rm e}\right)}{4\zeta_{\rm e}}
\nonumber \\
&&+\tilde{\tau}\frac{\left(-1+\rho_{\rm e}\cos\theta_{\rm e}+\zeta_{\rm e}\right)}{2\rho_{\rm e}\sin\theta_{\rm e}},
\label{eq:18L}
\end{eqnarray}
where $\rho_{\rm e}$ and $\zeta_{\rm e}$ correspond to those when $\theta=\theta_{\rm e}$ determined from the generalized Young's equation (\ref{eq:11L}).   Equation~(\ref{eq:18L}) reduces to the well-known formula~\cite{Navascues1981} for a droplet on a flat substrate when $\rho\rightarrow 0$ or $R\rightarrow\infty$.

\begin{figure}[htbp]
\begin{center}
\includegraphics[width=0.70\linewidth]{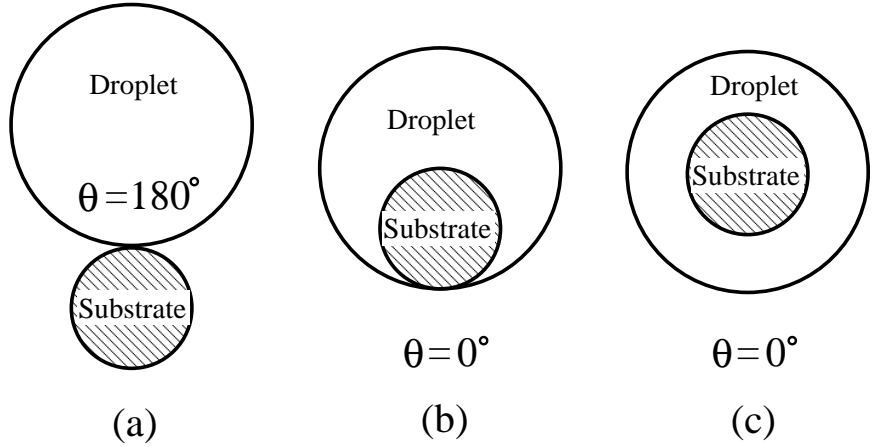}
\caption{
(a) A spherical droplet sitting on top of a spherical substrate. The contact angle is $\theta=180^{\circ}$.  (b) A spherical droplet wrapped around a spherical substrate.  The contact angle is $\theta=0^{\circ}$.  (c) A spherical droplet with a boiled-egg-like structure; it has the same free energy as that in part (b).  Therefore, this structure is also one of the structures which has the same free energy as that for $\theta=0^{\circ}$.  }
\label{fig:3L}
\end{center}
\end{figure}

On the other hand, the free energy of a spherical droplet sitting on top of a spherical substrate [Fig.~\ref{fig:3L}(a)] is given by Eq.~(\ref{eq:3L}) when $\theta=180^{\circ}$ and is written as 
\begin{equation}
F_{\rm sphere}=4\pi R^{2}\sigma_{\rm lv}f_{180}
\label{eq:19L}
\end{equation}
where
\begin{equation}
f_{180}=\rho_{180}^{2}
\label{eq:20L}
\end{equation}
and $\rho_{180}$ is the size parameter when the contact angle is $\theta=180^{\circ}$.  It is also possible to calculate the free energy of a droplet that completely wraps the spherical substrate~\cite{Kuni1996,Bieker1998,Bykov2006} [Fig.~\ref{fig:3L}(b)], which is realized when $\rho>1$ and $\theta=0$.  The free energy is given again by Eq.~(\ref{eq:3L}) when $\theta=0^{\circ}$: 
\begin{equation}
F_{\rm wrap}=4\pi R^{2}\sigma_{\rm lv}f_{0}
\label{eq:21L}
\end{equation}
where
\begin{equation}
f_{0}=\rho_{0}^{2}-\cos\theta_{\rm Y}
\label{eq:22L}
\end{equation}
and $\rho_{0}$ is the size parameter when the contact angle is $\theta=0^{\circ}$.  This free energy is the same as that of the boiled-egg-like structure, shown in Fig.~\ref{fig:3L}(c).  By comparing the free energy $f_{\rm e}$ of a cap-shaped droplet in Eq.~(\ref{eq:18L}) with $f_{180}$ of a detached spherical droplet in Eq.~(\ref{eq:20L}), we can study the drying transition~\cite{Widom1995} on a spherical substrate.  Similarly, by comparing the free energy $f_{\rm e}$ of a cap-shaped droplet with $f_{0}$ of a wrapped spherical droplet in Eq.~(\ref{eq:22L}), we can determine the wetting transition on a spherical substrate.

\begin{figure}[htbp]
\begin{center}
\includegraphics[width=0.70\linewidth]{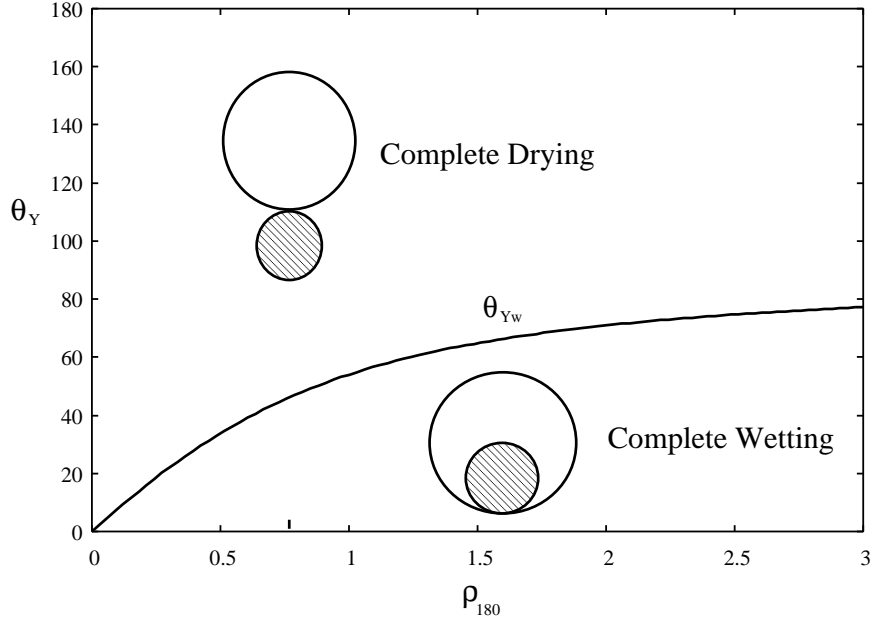}
\caption{
The wetting-drying boundary $\theta_{\rm Yw}$ of detached and wrapped spherical droplets, defined by Eq.~(\ref{eq:24L}) as a function of the radius (volume) $\rho_{180}$ of the droplet.  Below  $\theta_{\rm Yw}$, a wrapped droplet is more stable than a detached droplet.   Therefore, if the substrate is hydrophilic ($\theta_{\rm Y}<\theta_{\rm Yw}<90^{\circ}$), the complete wetting of the wrapped droplet is favorable.  
}
\label{fig:4L}
\end{center}
\end{figure}

The transition line separating a wrapped spherical droplet that encloses a spherical substrate from a detached, spherical droplet on top of a spherical substrate is determined by $F_{\rm sphere}=F_{\rm wrap}$ in Eqs.~(\ref{eq:19L}) and (\ref{eq:21L})  together with the conservation of the droplet volume
\begin{equation}
\rho_{0}^3-1=\rho_{180}^{3},
\label{eq:23L}
\end{equation}
which leads to the condition of the wetting-drying boundary
\begin{equation}
\theta_{\rm Yw}=\cos^{-1}\left(\left(\rho_{180}^{3}+1\right)^{2/3}-\rho_{180}^{2}\right)
\label{eq:24L}
\end{equation}
for the Young's contact  angle.  Figure~\ref{fig:4L} shows the wetting-drying boundary $\theta_{\rm Yw}$ as a function of the radius (volume) $\rho_{180}$ of the droplet.  Below this curve, the substrate is hydrophilic, and the completely wetted configuration of the wrapped droplet has a lower free energy than the completely dried configuration of the detached droplet.  Along this line $\theta_{\rm Yw}$, a wrapped spherical droplet and a detached spherical droplet have the same free energy. 

We analyze the transformation of a cap-shaped droplet into a detached, spherical morphology, which is realized when $F_{\rm sphere}=F_{\rm cap}$, in the following section.  We refer to this transition as the complete-drying transition or simply the drying transition, though the drying transition of the surface-phase transition is used for the open system~\cite{Dietrich1988,Bonn2009}, so both the volume of the droplet and the number of molecules can change.  Similarly, we refer to the morphological transition from cap-shaped to wrapped sphere, which is realized when $F_{\rm wrap}=F_{\rm cap}$, as the complete-wetting transition or simply the wetting transition.  This transition is not the surface-induced phase transition~\cite{Dietrich1988,Bonn2009} but is the morphological transition under the condition of a fixed number of molecules or fixed volume.

\section{\label{sec:sec3}  Morphological transition of a cap-shaped droplet on a spherical substrate}

\begin{figure}[htbp]
\begin{center}
\includegraphics[width=0.70\linewidth]{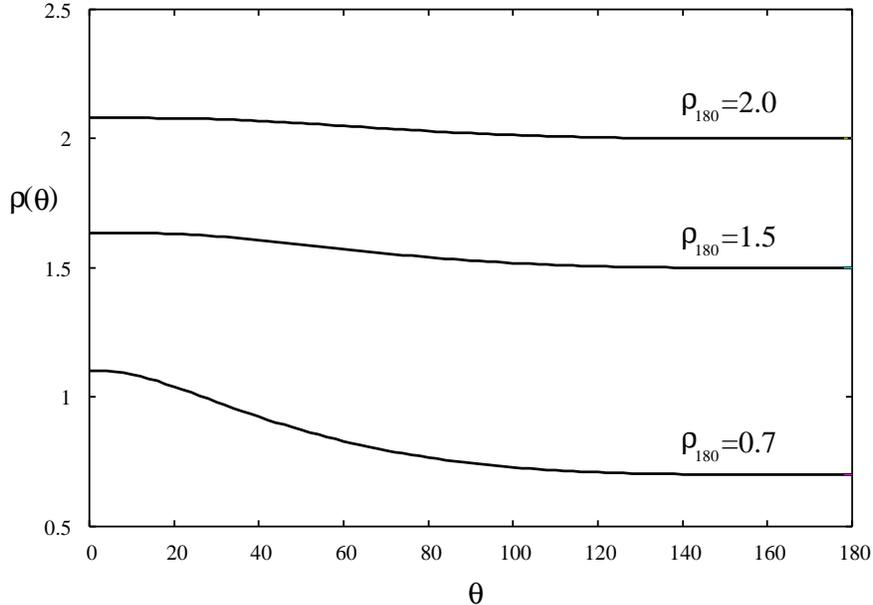}
\caption{
The size parameter $\rho\left(\theta\right)$ as a function of the contact angle $\theta$ for $\rho_{180}=0.7, 1.5, {\rm and}\;2.0$.  The scaled radius $\rho$ of the droplet increases as the contact angle $\theta$ decreases from $\rho_{180}$ for $\theta=180^{\circ}$.   
}
\label{fig:5L}
\end{center}
\end{figure}

Since the droplet volume is conserved, the radius $r$ of the droplet changes as the contact angle $\theta$ on the spherical substrate is altered.   For a given spherical substrate with radius $R$, the droplet volume characterized by $\rho_{180}$, and the size parameter $\rho\left(\theta\right)$ as a function of the contact angle $\theta$, are determined from the implicit equation  
\begin{equation}
\omega\left(\rho,\theta\right)=\rho_{180}^{3}.
\label{eq:25L}
\end{equation}
Figure \ref{fig:5L} shows the size parameter $\rho\left(\theta\right)$ as a function of the contact angle $\theta$ for a given value of $\rho_{180}$.  The size parameter $\rho$ and, therefore, the radius $r$ of the droplet increases from $\rho_{180}$ at $\theta=180^{\circ}$ as the contact angle decreases.  The three-phase contact line  coincides with the equator when the contact angle becomes the characteristic contact angle ($\theta=\theta_{\rm c}$), determined from Eq.~(\ref{eq:15L}) even when $\rho_{180}=0.7$, because $\rho$ can be larger than 1 for small contact angles.

\begin{figure}[htbp]
\begin{center}
\includegraphics[width=0.70\linewidth]{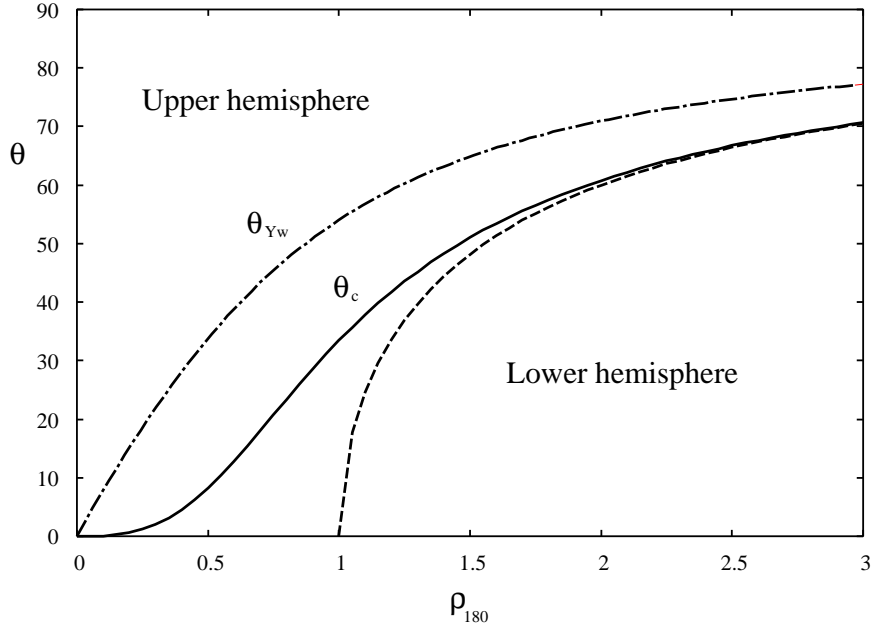}
\caption{
The characteristic contact angle $\theta_{\rm c}$ as a function of the size parameter $\rho_{180}$ (solid curve). We also show the contact angle $\theta=\cos^{-1}\left(1/\rho_{180}\right)$ for the fixed radius $\rho_{180}$ (dashed curve), which is characteristic of the nucleation problem~\cite{Iwamatsu2015,Iwamatsu2015b}.  The contact line can cross the equator as the size parameter $\rho\left(\theta\right)$ changes as a function of $\theta$ (see Fig.~\ref{fig:4L}) when the droplet volume is fixed by $\rho_{180}^{3}$. If the scaled radius of the droplet is fixed, the droplet cannot reach the equator if $\rho_{180}<1$, and no characteristic contact angle $\theta_{\rm c}$ can be found as a solution of Eq.~(\ref{eq:15L}).  We also show the wetting-drying boundary (dash-dotted curve) depicted in Fig.~\ref{fig:4L}, which is always larger than $\theta_{\rm c}$.
 }
\label{fig:6L}
\end{center}
\end{figure}

Figure \ref{fig:6L} shows the characteristic contact angle $\theta_{\rm c}$ calculated from Eq.~(\ref{eq:15L}).  Even when the scaled radius $\rho_{180}$ for a spherical droplet is smaller than 1 ($\rho_{180}<1$), a characteristic contact angle $\theta_{\rm c}$ exists when the droplet volume is fixed because the scaled radius $\rho_{\rm c}$ in Eq.~(\ref{eq:15L}) can be larger than 1 (Fig.~\ref{fig:5L}) and Eq.~(\ref{eq:15L}) has a solution.  Therefore, the contact line of a droplet placed on top of a spherical substrate can cross the equator and can reach the lower hemisphere even when $\rho_{180}<1$.  On the other hand, when the radius $r$ rather than the volume of the droplet is fixed and $\rho_{180}=r/R$, which occurs in the nucleation problem~\cite{Iwamatsu2015,Iwamatsu2015b}, the characteristic contact angle $\theta_{\rm c}$ does not exist and the contact line cannot reach the equator when $\rho_{180}<1$ because Eq.~(\ref{eq:15L}) has no solution for $\rho_{180}=\rho_{\rm c}<1$.

Since the droplet radius $\rho$ is a function of the contact angle $\theta$ from the implicit equation Eq.~(\ref{eq:25L}), the free energy Eq.~(\ref{eq:4L}) becomes a function of the contact angle $\theta$.  Then, the morphology of the droplet on a spherical substrate is determined by the absolute minimum of the free energy landscape $f\left(\theta\right)=f\left(\rho\left(\theta\right), \theta\right)$.

\begin{figure}[htbp]
\begin{center}
\includegraphics[width=0.70\linewidth]{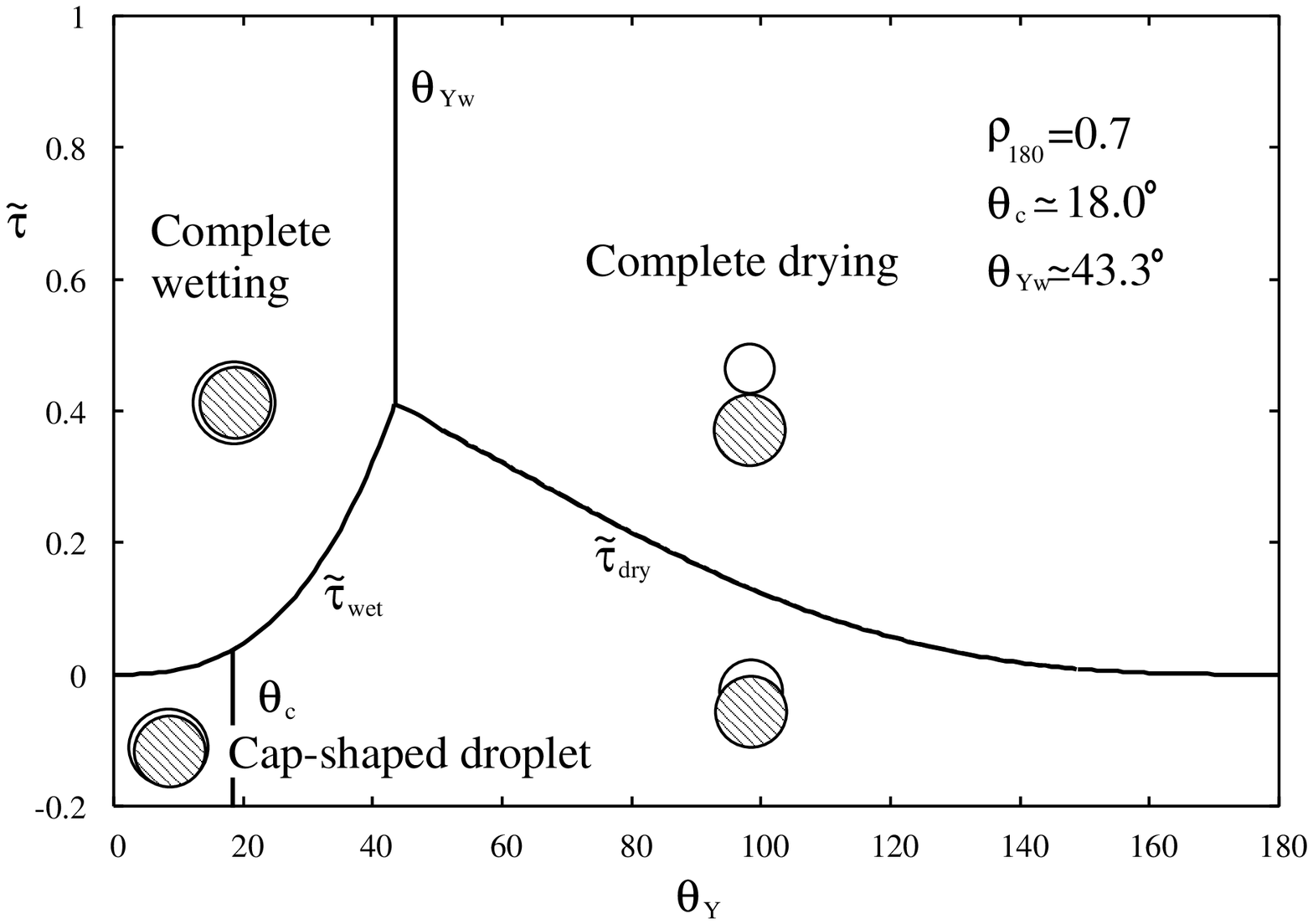}
\caption{
Morphological phase diagram of the droplet on a spherical substrate when $\rho_{180}=0.7$.  There are four regions separated by two solid curves and two vertical lines.  These four regions corresponds to a wrapped spherical droplet (complete wetting, $\theta=0^{\circ}$), a detached spherical droplet (complete drying, $\theta=180^{\circ}$), a cap-shaped droplet ($\theta>\theta_{\rm c}$) whose contact line is on the upper hemisphere of the substrate, and the one ($\theta<\theta_{\rm c}$) whose contact line is on the lower hemisphere.  
}
\label{fig:7L}
\end{center}
\end{figure}

Figure~\ref{fig:7L} shows the phase diagram of morphological transitions between a detached spherical droplet (complete drying), a wrapped spherical droplet (complete wetting), a cap-shaped droplet whose contact line is on the upper hemisphere of the substrate ($\theta>\theta_{\rm c}$), and a cap shaped droplet whose contact line is on the lower hemisphere ($\theta<\theta_{\rm c}$) when $\rho_{180}=0.7$.  The diagram is divided into four regions where these four morphologies have the lowest free energy. Therefore, the lines indicate the locus of transitions where the global minimum of the free energy changes.

The transition between a detached droplet and a wrapped droplet is given by Eq.~(\ref{eq:24L}) and is represented by the vertical line at $\theta=\theta_{\rm Yw}$.  The transition between a detached spherical droplet and a cap-shaped droplet (complete drying transition) is given by the curve indicated by $\tilde{\tau}_{\rm dry}$, which is defined by the condition $f_{\rm e}=f_{180}$ given by Eqs.~(\ref{eq:18L}) and (\ref{eq:20L}).  The transition between a wrapped spherical droplet and a cap-shaped droplet (complete wetting transition) is given by the curve indicated by $\tilde{\tau}_{\rm wet}$, which is defined by the condition $f_{\rm e}=f_{0}$ given by Eqs.~(\ref{eq:18L}) and (\ref{eq:22L}).  

Even beyond these transition curves $\tilde{\tau}_{\rm dry}$ and $\tilde{\tau}_{\rm wet}$, the cap-shaped droplet can exist as a metastable droplet whose free energy is at the local minimum rather than at the global (lowest) minimum.  The stability limit of this metastable cap-shaped droplet is similar to the {\it spinodal} line of a first-order phase transition, which can be determined from the second derivative of the free energy, as shown in Appendix.  This stability limit is given by
\begin{equation}
\tilde{\tau}_{\rm st}\left(\theta_{\rm e}\right)
=\frac{\rho_{\rm e}\left(-\rho_{\rm e}+\cos\theta_{\rm e}+2\sqrt{1+\rho_{\rm e}^{2}-2\rho_{\rm e}\cos\theta_{\rm e}}\right)\sin^{3}\theta_{\rm e}}{1+\rho_{\rm e}^{2}-2\rho_{\rm e}\cos\theta_{\rm e}},
\label{eq:26L}
\end{equation}
which becomes
\begin{equation}
\tilde{\tau}_{\rm st}\left(\theta_{\rm e}\right) \rightarrow \sin^{3}\left(\theta_{\rm e}\right),\;\;\;\mbox{as}\;\;\;\;\;\rho_{\rm e}\rightarrow \infty,
\label{eq:27L}
\end{equation}
where $\theta_{\rm e}$ is the equilibrium contact angle determined from Eq.~(\ref{eq:11L}) for the given Young's contact angle $\theta_{\rm Y}$ and the fixed droplet volume. When the line tension is larger than $\tilde{\tau}_{\rm st}$ ($\tilde{\tau}>\tilde{\tau}_{\rm st}$), the cap-shaped droplet maximizes the Helmholtz free energy, and does not represent a mechanically stable solution to the first
variational problem. Then, only a detached spherical droplet sitting on a spherical substrate, as shown in Fig.~\ref{fig:3L}(a), forms when $\theta_{\rm Y}>\theta_{\rm Yw}$ or the wrapped spherical droplet, as shown in Figs.~\ref{fig:3L}(b) and \ref{fig:3L}(c), forms when $\theta_{\rm Y}<\theta_{\rm Yw}$ according to Fig.~\ref{fig:4L}.

\begin{figure}[htbp]
\begin{center}
\includegraphics[width=0.70\linewidth]{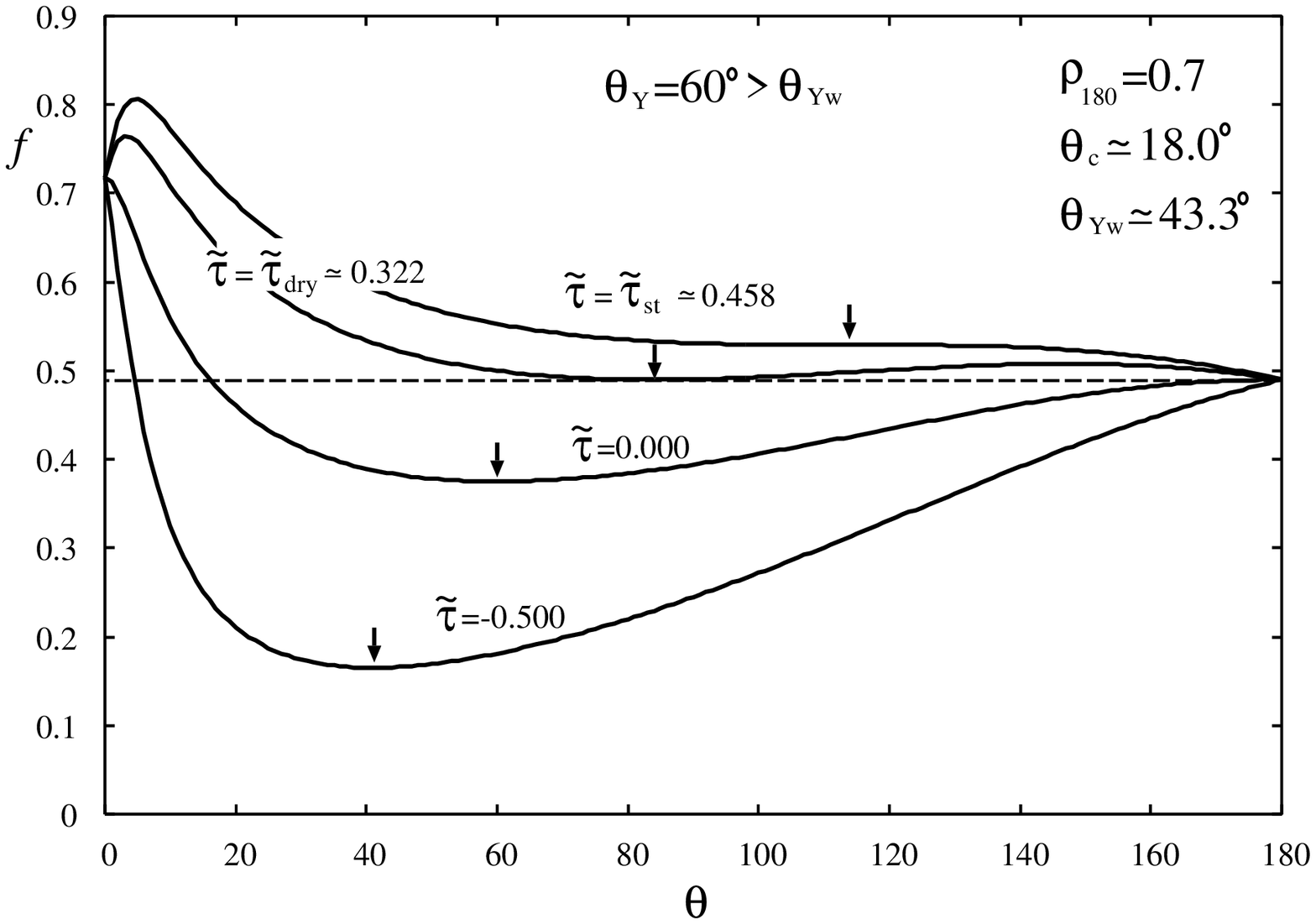}
\caption{
The free-energy landscape $f$ given by Eq.~(\ref{eq:4L}) as a function of the contact angle $\theta$ for $\rho_{180}=0.7$ when $\theta_{\rm Y}=60^{\circ}>\theta_{\rm Yw}$. The local minimum at $\theta=0^{\circ}$ corresponds to the wrapped spherical droplet (complete wetting) and that at $\theta=180^{\circ}$ corresponds to the detached spherical droplet (complete drying).  The local minimum indicated by short arrows correspond to the stable and the metastable cap-shaped droplets whose contact angle $\theta_{\rm e}$ is determined from Eq.~(\ref{eq:11L}).  The arrows on the stability limit $\tilde{\tau}_{\rm st}$ indicates the contact angle $\theta_{\rm e}$ of the unstable cap-shaped droplet determined from Eq.~(\ref{eq:11L}).  The stable structure with the minimum free energy is a spherical droplet with $\theta=180^{\circ}$ when $\tilde{\tau}>\tilde{\tau}_{\rm dry}\simeq 0.322$.  The metastable cap-shaped droplet becomes unstable when $\tilde{\tau}>\tilde{\tau}_{\rm st}\simeq 0.458$.   On the other hand, the cap-shaped droplet is stable and the equilibrium contact angle $\theta_{\rm e}$ approaches the characteristic contact angle $\theta_{\rm c}\simeq 18.0^{\circ}$ when the line tension is large negative.
}
\label{fig:8L}
\end{center}
\end{figure}

Figure~\ref{fig:8L} shows the free-energy landscape when $\theta_{\rm Y}=60^{\circ}>\theta_{\rm Yw}$.  The landscape shows a minimum near the characteristic contact angle $\theta_{\rm e}=41.0^{\circ}>\theta_{\rm c}\simeq 18.0^{\circ}$ when $\tilde{\tau}=-0.5$.  In fact, the equilibrium contact angle $\theta_{\rm e}$ indicated by a short arrow is determined from Eq.~(\ref{eq:11L}).  It approaches $\theta_{\rm c}\simeq18.0^{\circ}$ from above since contact line approaches the equator from the upper hemisphere to maximize the contact-line length when the line tension is large negative.  When the line tension vanishes ($\tilde{\tau}=0$), the equilibrium contact angle is given by the Young's contact angle $\theta_{\rm e}=\theta_{\rm Y}=60^{\circ}$.  

Further increase in the positive line tension leads to the retraction of the contact line towards the north pole of the spherical substrate where the droplet is located, and to the increase in the equilibrium contact angle $\theta_{\rm e}$.  When the line tension reaches $\tilde{\tau}_{\rm dry}\simeq 0.322$ of the drying transition, the free energy of the cap-shaped droplet at $\theta_{\rm e}=83.8^{\circ}$ becomes equal to that of the detached spherical droplet at $\theta=180^{\circ}$.  Then the cap-shaped droplet can transform into the detached spherical droplet sitting on the top of a spherical substrate.  However, this transition is discontinuous and thus exhibit a free energy barrier, which need to be overcome by some fluctuations.  Above $\tilde{\tau}_{\rm dry}$, the cap-shaped droplet can exist as a metastable object, which becomes unstable at the stability limit $\tilde{\tau}_{\rm st}\simeq 0.458$.  This point is similar to the {\it spinodal} of the first-order phase transition.

\begin{figure}[htbp]
\begin{center}
\includegraphics[width=0.70\linewidth]{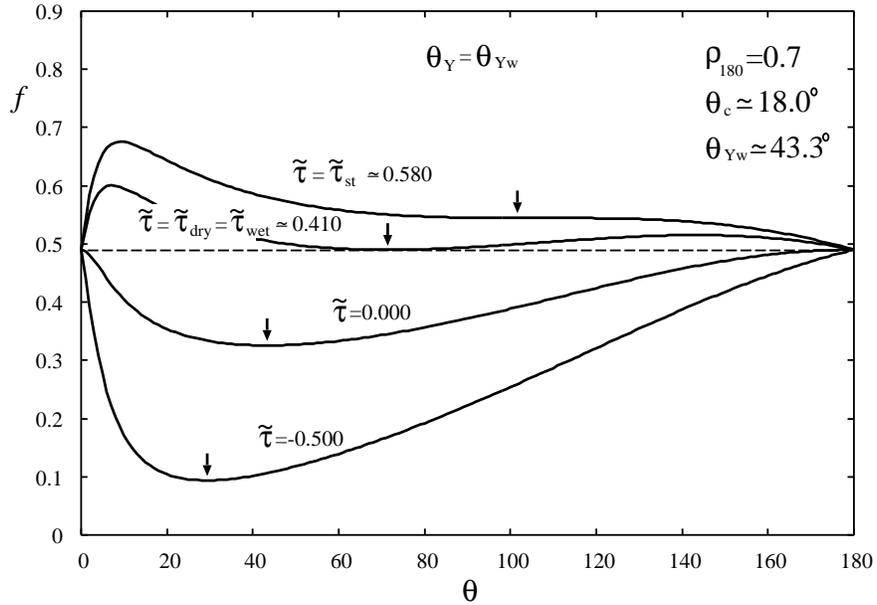}
\caption{
The free-energy landscape $f$ as a function of the contact angle $\theta$ for $\rho_{180}=0.7$ when $\theta_{\rm Y}=\theta_{\rm Yw}\simeq 43.3^{\circ}$.   In this special case of the hydrophilic-hydrophobic boundary, both the drying transition and the wetting transition can occur.  When $\tilde{\tau}=\tilde{\tau}_{\rm dry}=\tilde{\tau}_{\rm wet}\simeq 0.410$, the landscape shows three minimums of equal depth at $\theta=0^{\circ}$, at $\theta=180^{\circ}$, and at $\theta_{\rm e}=71.5^{\circ}$, where the cap-shaped droplet may transform into either the detached spherical droplet or the wrapped spherical droplet.
}
\label{fig:9L}
\end{center}
\end{figure}

Figure~\ref{fig:9L} shows the free-energy landscape when $\theta_{\rm Y}=\theta_{\rm Yw}\simeq 43.3^{\circ}$.  In this special case, the drying transition and the wetting transition coexist.  When $\tilde{\tau}=\tilde{\tau}_{\rm dry}=\tilde{\tau}_{\rm wet}\simeq 0.410$, the landscape shows three minima of equal depth at $\theta=0^{\circ}$ of wrapped spherical droplet (complete wetting), at $\theta=180^{\circ}$ of the detached spherical droplet (complete drying), and at $\theta_{\rm e}=71.5^{\circ}$ of the cap-shaped droplet.  Then, the cap-shaped droplet may transform into either the wrapped spherical droplet or the detached spherical droplet.  Again, the metastable cap-shaped droplet becomes unstable at $\tilde{\tau}=\tilde{\tau}_{\rm st}\simeq 0.580$.

\begin{figure}[htbp]
\begin{center}
\includegraphics[width=0.70\linewidth]{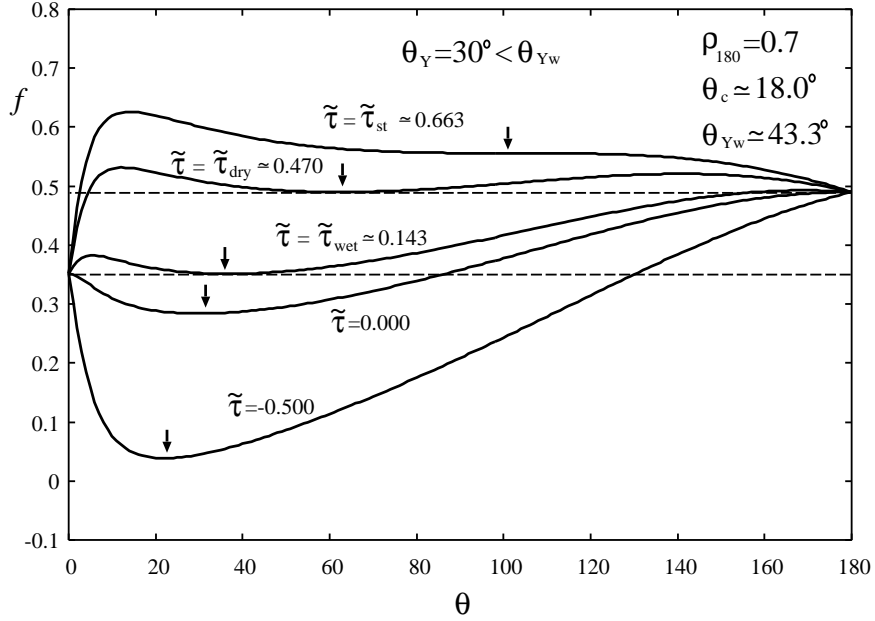}
\caption{
The free-energy landscape $f$ as a function of the contact angle $\theta$ for $\rho_{180}=0.7$ when $\theta_{\rm Y}=30^{\circ}<\theta_{\rm Yw}\simeq 43.3^{\circ}$.   In this case, the substrate is hydrophilic. Therefore, the wetting transition occurs when $\tilde{\tau}=\tilde{\tau}_{\rm wet}\simeq 0.143$.   However, a pseudo drying transition from the metastable cap-shaped droplet to the metastable detached spherical droplet occurs when $\tilde{\tau}=\tilde{\tau}_{\rm dry}\simeq 0.470$. 
}
\label{fig:10L}
\end{center}
\end{figure}

When $\theta_{\rm Y}<\theta_{\rm Yw}$, the cap-shaped droplet transforms into the wrapped spherical droplet and the complete wetting rather than the complete drying transition occurs.  Figure~\ref{fig:10L} shows the free-energy landscape when $\theta_{\rm Y}=30^{\circ}$.  The contact angle increases as the line tension is increased.  The free energy of the cap-shaped droplet with the contact angle $\theta_{\rm e}=36.4^{\circ}$ becomes equal to that of the wrapped spherical droplet with the contact angle $0^{\circ}$ when $\tilde{\tau}=\tilde{\tau}_{\rm wet}\simeq 0.143$, where the complete wetting transition occur. By further increasing the magnitude of the line tension, the free energy of the metastable cap-shaped droplet with the contact angle $\theta_{\rm e}=62.1^{\circ}$ becomes equal to that of the detached spherical droplet with $\theta=180^{\circ}$ at $\tilde{\tau}=\tilde{\tau}_{\rm dry}\simeq 0.470$.  Then, the metastable cap-shaped droplet may transform into the metastable detached spherical droplet (pseudo drying transition).  The metastable cap-shaped droplet  becomes unstable at $\tilde{\tau}=\tilde{\tau}_{\rm st}\simeq 0.580$.

\begin{figure}[htbp]
\begin{center}
\includegraphics[width=0.70\linewidth]{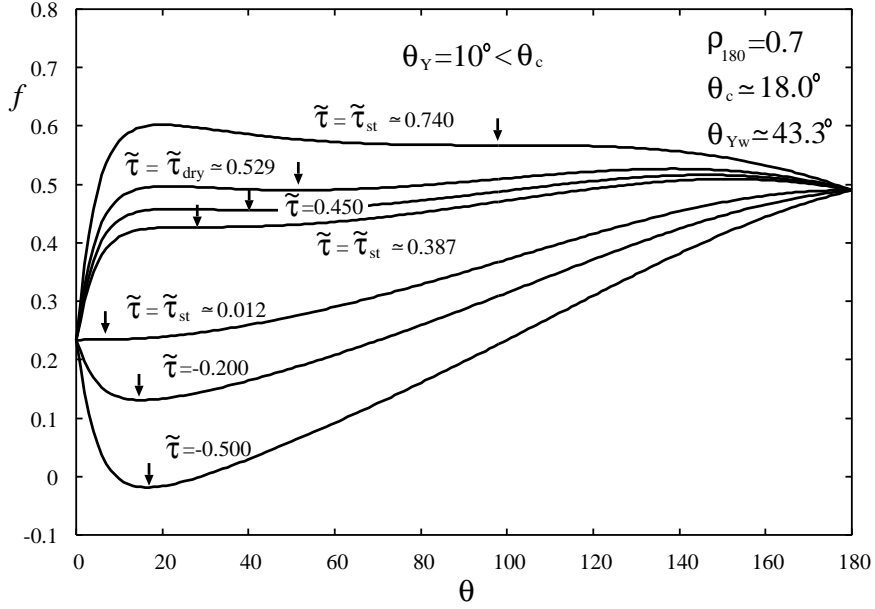}
\caption{
The free-energy landscape $f$ as a function of the contact angle $\theta$ for $\rho_{180}=0.7$ when $\theta_{\rm Y}=10^{\circ}<\theta_{\rm c}\simeq 18.0^{\circ}$.  In this case, not only the cap-shaped droplet with a contact line on the lower hemisphere but also the one with a contact line on the upper hemisphere can exist.   There are three stability limits $\tilde{\tau}_{\rm st}$.  At the first stability limit $\tilde{\tau}=\tilde{\tau}_{\rm st}\simeq 0.012$, the cap-shaped droplet  with a contact line on the lower hemisphere ($\theta<\theta_{\rm c}\simeq 18.0^{\circ}$) becomes unstable.  At the next stability limit $\tilde{\tau}=\tilde{\tau}_{\rm st}\simeq 0.387$, the metastable cap-shaped droplet with a contact line on the upper hemisphere ($\theta>\theta_{\rm c}\simeq 18.0^{\circ}$) reappear.  This cap-shaped droplet  with a contact line on the upper hemisphere may transform into the metastable detached sphere at $\tilde{\tau}_{\rm dry}\simeq 0.529$.  Finally, this metastable cap-shaped droplet with a contact line on the upper hemisphere becomes unstable at the third stability limit $\tilde{\tau}_{\rm st}\simeq 0.740$.
}
\label{fig:11L}
\end{center}
\end{figure}

When $\theta_{\rm Y}<\theta_{\rm c}\simeq 18.0^{\circ}$, a stable cap-shaped droplet  with a contact line only on the lower hemisphere can exist.  However, both the metastable droplet  with a contact line on the upper hemisphere ($\theta >\theta_{\rm c}$) and that on the lower hemisphere ($\theta< \theta_{\rm c}$) can exist.   Figure~\ref{fig:11L} shows the free-energy landscape when $\theta_{\rm Y}=10^{\circ}<\theta_{\rm c}$.  The free energy of the cap-shaped droplet becomes equal to that of the wrapped spherical droplet of the contact angle $0^{\circ}$ when $\tilde{\tau}=\tilde{\tau}_{\rm wet}\simeq 0.0068$, where the complete wetting transition, which is not shown in Fig.~\ref{fig:11L}, occurs.  

By further increasing the magnitude of the line tension, the metastable cap-shaped droplet with a contact line on the lower hemisphere soon becomes unstable at the first stability limit $\tilde{\tau}=\tilde{\tau}_{\rm st}\simeq 0.012$.  After this first stability limit, the metastable cap-shaped droplet cannot exist until the line tension reaches the second stability limit $\tilde{\tau}=\tilde{\tau}_{\rm st}\simeq 0.387$.  The metastable cap-shaped droplet reappears after this second stability limit, now, with a contact line on the upper hemisphere.  Then, the metastable cap-shaped droplet with a contact line on the upper hemisphere may transform into a detached metastable spherical droplet (pseudo drying transition) at $\tilde{\tau}=\tilde{\tau}_{\rm dry}\simeq 0.529$.  This metastable cap-shaped droplet with a contact line on the upper hemisphere becomes finally unstable at the third stability limit $\tilde{\tau}=\tilde{\tau}_{\rm st}\simeq 0.740$.

\begin{figure}[htbp]
\begin{center}
\includegraphics[width=0.70\linewidth]{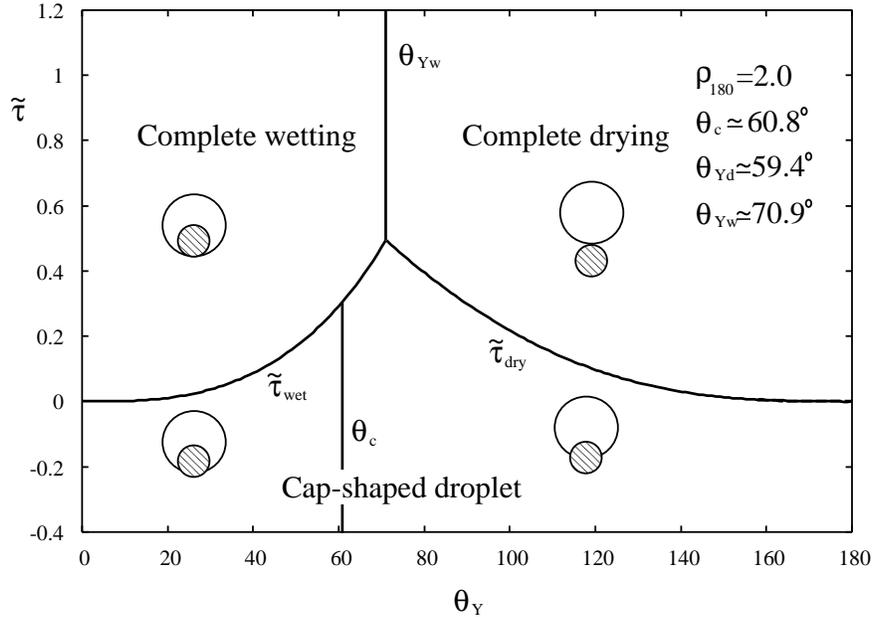}
\caption{
The morphological phase diagram of the droplet on a spherical substrate when $\rho_{180}=2.0$, which also shows four regions.  
}
\label{fig:12L}
\end{center}
\end{figure}

Figure~\ref{fig:12L} shows the phase diagram of morphological transition when $\rho_{180}=2.0$.  The droplet volume is larger than that of the spherical substrate.  The wetting-drying boundary $\theta_{\rm Yw}$ and the characteristic contact angle $\theta_{\rm c}$ shift to higher angles $\theta_{\rm Yw}\simeq 70.9^{\circ}$ and $\theta_{\rm c}\simeq 60.8^{\circ}$.  

\begin{figure}[htbp]
\begin{center}
\includegraphics[width=0.70\linewidth]{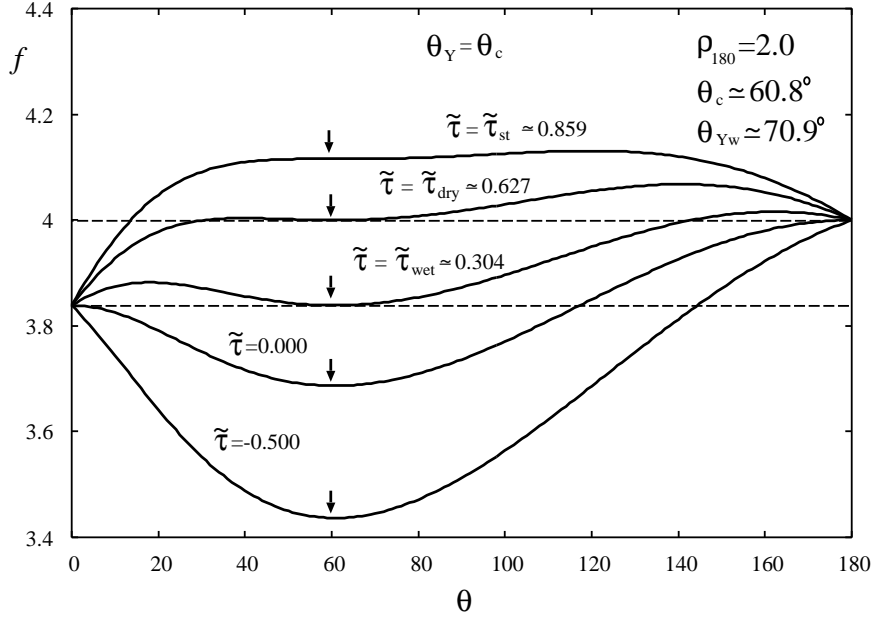}
\caption{
The free-energy landscape $f$ as a function of the contact angle $\theta$ for $\rho_{180}=2.0$ when $\theta_{\rm Y}=\theta_{\rm c}\simeq 60.8^{\circ}$.  In this case, the contact angle is fixed at $\theta=\theta_{\rm c}$, which means that the contact line of the cap-shaped droplet cannot move and it is fixed at the equator of the spherical substrate.
}
\label{fig:13L}
\end{center}
\end{figure}

Figure~\ref{fig:13L} shows the free-energy landscape when $\theta_{\rm Y}=\theta_{\rm c}\simeq 60.8^{\circ}$.  The landscape shows a minimum at the equilibrium contact angle $\theta_{\rm e}$, which is exactly at the characteristic contact angle $\theta_{\rm e}=\theta_{\rm c}\simeq 60.8^{\circ}$ as discussed in Eq.~(\ref{eq:15L}).  Therefore, the three-phase contact line of the cap shaped droplet is fixed at the equator of the spherical substrate.  The cap-shaped droplet may transform into the wrapped spherical droplet at $\tilde{\tau}_{\rm wet}\simeq 0.304$ (complete wetting transition).  By further increasing the line tension, the metastable cap-shaped droplet may transform into a detached metastable spherical droplet (pseudo drying transition) at $\tilde{\tau}=\tilde{\tau}_{\rm dry}\simeq 0.627$.  The metastable cap-shaped droplet whose contact line is fixed at the equator becomes finally unstable at $\tilde{\tau}=\tilde{\tau}_{\rm st}\simeq 0.859$.

The effects of negative line tension on the droplet on a spherical substrate are different from those on a flat substrate.  As we increase the magnitude of the negative line tension ($\tilde{\tau}<0$), the contact angle asymptotically approaches the characteristic contact angle $\theta_{\rm c}$ from above ($\theta\rightarrow\theta_{\rm c}^{+}$) when the three-phase contact line is located on the upper hemisphere, and it approaches $\theta_{\rm c}$ from below ($\theta\rightarrow\theta_{\rm c}^{-}$) when the three-phase contact line is located on the lower hemisphere.  Therefore, the three-phase contact line indefinitely approaches the equator of the substrate from both above and below.  However, the contact line always remains in its original hemisphere, and cannot cross the equator.  This finding can be easily understood, because the contact-line length is maximized at the equator.  In order to maximize the negative gain of the line-tension contribution of the Helmholtz free energy, the contact line approaches the equator but never crosses it.  Therefore, the droplet always remains on its original hemisphere, irrespective of the magnitude of the line tension.  

Since this study is concerned with the thermodynamics of a droplet, the stability of a cap-shaped droplet against fluctuations that do not preserve its circular shape is not considered.  It is well known that the capillary model of the cap-shaped droplet employed in this work possesses short-wavelength instability~\cite{Dobbs1999,Brinkmann2005,Guzzardi2006} for negative line tension on a flat substrate since the undulation of the contact line around the circular shape will necessarily increase the contact-line length and will decrease the free energy.  However, Mechkov et al.~\cite{Mechkov2007} pointed out that this instability is not physical when the molecular interaction near the three-phase contact line is included using the disjoining pressure and the interface-displacement model~\cite{Indekeu1992}.  We will leave this problem of the fluctuation and the inclusion of the disjoining pressure on a spherical substrate for future investigation.

\section{\label{sec:sec5}Conclusion}

In this study, we considered the line-tension effects on the stability of a cap-shaped droplet of a fixed volume, placed on a spherical substrate. We found that the contact angle is determined by the generalized Young's equation, as it takes into account the effects of the line tension.  The analytical expression for the Helmholtz free energy is found, and it consists of the usual surface contribution and a contribution from the line tension.   

Using the generalized Young's equation, We studied the contact angle of a cap-shaped droplet as a function of the line tension. The morphological transition from a cap-shaped droplet to a detached, spherical droplet and a wrapped, spherical droplet was examined by comparing the minimized Helmholtz free energy of a cap-shaped droplet with that of spherical droplets of the same volume. 

We found a special role played by the equator of the spherical substrate, at which the contact-line length of a droplet reached its maximum.  The contact line of the droplet cannot cross the equator while changing the magnitude of line tension continuously.  When the contact line of a droplet coincided with the equator, the droplet with this characteristic contact angle is pinned at the equator for a positive line tension unless its magnitude does not exceeds that for the wetting transition.  The droplet is also pinned for a large negative line tension.    

When the Young's contact angle is larger than the wetting-drying boundary ($\theta_{\rm Y}>\theta_{\rm Yw}$), upon increasing the {\it positive line tension}, the Helmholtz free energy of a cap-shaped droplet could exceed that of a detached, spherical droplet.  Then, the contact angle jumps from a finite value to $180^{\circ}$.  This morphological transition is the same as the {\it complete-drying transition} predicted by Widom~\cite{Widom1995} for a droplet on a flat substrate.  

On the other hand, when  the Young's contact angle is smaller than the wetting-drying boundary ($\theta_{\rm Y}<\theta_{\rm Yw}$), upon increasing the {\it positive line tension}, the Helmholtz free energy of a cap-shaped droplet could exceed that of a wrapped, spherical droplet.  Then, the contact angle jumps from a finite value to $0^{\circ}$.  This morphological transition is similar to the {\it complete-wetting transition}~\cite{Bonn2009, Dietrich1988} of the surface phase transition.

On increasing the absolute magnitude of the {\it negative line tension}, the contact angle approaches the characteristic contact angle, and, therefore, the contact line approaches the equator asymptotically from either above or below. However, this result is not conclusive as the undulation of the contact line necessarily decreases the free energy and a circular contact line might be marginally stable.  In fact, the interfacial shape becomes unstable for a sufficiently large negative line tension.  This problem is know to be partly due to the shortcomings of the present capillary model, and it is resolved by introducing the disjoining pressure using the interface-displacement model~\cite{Mechkov2007}.  Similarly, the unphysical result of our capillary model in the $\theta\rightarrow 0^{\circ}$ limit, noted in our previous papers~\cite{Iwamatsu2015,Iwamatsu2015b}, can also be avoided by introducing the disjoining pressure, which allows for a preexisting thin liquid film.  The problem of the fluctuation of the contact line and the effects of the disjoining pressure remain for future investigation.

In this study, we analyzed various scenarios concerning a cap-shaped droplet on a spherical substrate using both Young's contact angle and the size of the droplet as the two independent parameters.  The former can be controlled by changing the material and surface chemistry of the substrate.  The latter can also be controlled by changing the volume of the non-volatile liquid.  Therefore, experimental confirmation and verification of our predictions of the morphological transitions of a cap-shaped droplet on a spherical substrate should be feasible.

\begin{acknowledgments}
This work was partly conducted while I am a visiting scientist with the Department of Physics, Tokyo Metropolitan University.  I am grateful to Professors H. Mori and Y. Okabe for their continuous support.
\end{acknowledgments}

\appendix*

\section{}

Here, we provide the mathematical derivation of Eqs.~(\ref{eq:4L}), ~(\ref{eq:6L}), (\ref{eq:18L}), and (\ref{eq:26L}) in the main text. A more detailed derivation of these formulas has already been provided in my previous publications~\cite{Iwamatsu2015,Iwamatsu2015b}. Briefly, the derivation is based on the integration scheme proposed by Hamaker~\cite{Hamaker1937} and the change of variable from the contact angle $\theta$ to the distance $C$ between the centers of the two spheres of the substrate and the droplet, as shown in Fig.~\ref{fig:1L}.  By using this simple variable $C$, we avoid the tedious and complicated transformation of the trigonometric functions.  Because all these equations are linear in $\Delta\sigma$, $\sigma_{\rm lv}$, and $\tau$, the manipulation is tedious but straightforward.  

Now, the calculation of the volume $V$ is~\cite{Hamaker1937}
\begin{equation}
V=\frac{\pi}{12}\left(R+r-C\right)^{2}\left(C^{2}-3\left(R-r\right)^{2}+2C\left(R+r\right)\right),
\label{eq:AL1}
\end{equation}
which can be easily transformed into Eq.~(\ref{eq:6L}).

Similarly, the Helmholtz free energy can be calculated using the expression for the surface area
\begin{equation}
A_{\rm lv}=\pi r\frac{\left(r+C\right)^{2}-R^{2}}{C}
\label{eq:AL2}
\end{equation}
and
\begin{equation}
A_{\rm sl}=\pi R\frac{r^{2}-\left(R-C\right)^{2}}{C}
\label{eq:AL3}
\end{equation}
as well as the contact-line length
\begin{equation}
L=\pi\frac{\sqrt{2C^{2}\left(R^{2}+r^{2}\right)^{2}-\left(R^{2}-r^{2}\right)^{2}-C^{4}}}{C}.
\label{eq:AL4}
\end{equation}

By introducing the above three formulas into Eq.~(\ref{eq:1L}), the Helmholtz free energy is given by
\begin{equation}
F=\sigma_{\rm lv}\pi r\frac{(r+C)^{2}-R^{2}}{C}+\Delta\sigma\pi R\frac{r^{2}-(R-C)^{2}}{C}+\tau\frac{2\pi rR\sin\theta}{C}
\label{eq:AL5}
\end{equation}
which is reduced to Eqs.~(\ref{eq:3L}) and (\ref{eq:4L}).  After minimizing the free energy under the condition of constant volume $V$, we arrive at the equation
\begin{equation}
\Delta\sigma=-\frac{\left(-C^{2}+R^{2}+r^{2}\right)^{2}}{2rR}-\frac{\left(C^{2}+R^{2}-r^{2}\right)\tau}{R\sqrt{2C^{2}\left(R^{2}+r^{2}\right)^{2}-\left(R^{2}-r^{2}\right)^{2}-C^{4}}}.
\label{eq:AL6}
\end{equation}
which reduces to the generalized Young's equation~\cite{Iwamatsu2015} of Eq.~(\ref{eq:11L}).  Then, using the generalized Young's equation, the minimized free energy of the cap-shaped droplet is given by
\begin{eqnarray}
F_{\rm cap}&=&-\frac{\pi\left(C-R+r\right)^{2}\left(C^{2}-2rC-\left(R+r\right)^{2}\right)\sigma_{\rm lv}}{2rC}
\nonumber \\
&-&\frac{2\pi\left(C-R-r\right)\left(C-R+r\right)\tau}{\sqrt{\left(C+R-r\right)\left(C-R+r\right)\left(R+r-C\right)\left(R+r+C\right)}} \nonumber \\
\label{eq:AL7}
\end{eqnarray}
which can be rewritten as Eq.~(\ref{eq:18L}).

The second derivative is rather lengthy
\begin{eqnarray}
\frac{d^{2}F}{dr^{2}}
&=&-\frac{\pi\sigma_{\rm lv}\left(C^{2}-R^{2}-4rC+r^{2}\right)\left(C-R+r\right)^{2}\left(C+R+r\right)^{2}}{rC\left(C-R-r\right)^{2}\left(C+R-r\right)^{2}}
\nonumber \\
&+& 
\frac{16\pi r^{2}R^{2}C\tau \left(C-R+r\right)\left(C+R+r\right)\tau}{\left(C-R-r\right)^{3}\left(C+R-r\right)^{3}\sqrt{\left(R+r-C\right)\left(C+R-r\right)\left(C-R+r\right)\left(C+R+r\right)}} \nonumber \\
\label{eq:AL8}
\end{eqnarray}
Using the condition $d^{2}F/dr^{2}=0$ and changing the variable from $C$ to $\theta$, we arrive at Eq.~(\ref{eq:26L}).


\begin{thebibliography}{99}
\bibitem{deGennes1985} P. G. de Gennes, Rev. Mod. Phys. {\bf 57}, 827 (1985).
\bibitem{Dietrich1988} S. Dietrich, in {\it Phase Transition and Critical Phenomena}, edited by C. Domb and J. L. Lebowitz (Academic, London, 1988), Vol. 12, pp. 2-218.
\bibitem{Bonn2009} D. Bonn, J. Eggers, J. Indekeu,  J. Meunier, and E. Rolley, Rev. Mod. Phys. {\bf 81}, 739 (2009).
\bibitem{Bormashenko2013} E. Yu. Bormashenko {\it Wetting of Real Surfaces}, De Gruyter, Berlin, 2013.
\bibitem{Pompe2000} T. Pompe and S. Herminghaus, Phys. Rev. Lett. {\bf 85}, 1930 (2000).
\bibitem{Wang2001} J. Y. Wang, S. Betelu, and B. M. Law, Phys. Rev. E {\bf 63}, 031601 (2001).
\bibitem{Checco2003} A. Checco, P. Guenoun, and J. Daillant, Phys. Rev. Lett. {\bf 91}, 186101 (2003). 
\bibitem{Schimmele2007} L. Schimmele, M. Napi\'orkowski and S. Dietrich, J. Chem. Phys. {\bf 127}, 164715 (2007).
\bibitem{Navascues1981} G. Navascu\'es and P. Tarazona, J. Chem. Phys. {\bf 75}, 2441 (1981).
\bibitem{Widom1995} B. Widom, J. Phys. Chem. {\bf 99}, 2803 (1995).
\bibitem{Singha2015} S. K. Singha, P. K. Das, and B. Maiti, J. Chem. Phys. {\bf 142}, 104706 (2015). 
\bibitem{Iwamatsu2015} M. Iwamatsu, Langmuir {\bf 31}, 3861 (2015).
\bibitem{Lipowsky2001} R. Lipowsky, Interface Science {\bf 9}, 105 (2001).
\bibitem{Blecua2006} P. Blecua, R. Lipowsky, and J. Kierfeld, Langmuir {\bf 22}, 11041 (2006). 
\bibitem{Nosonovsky2007} M. Nosonovsky and B. Bhushan, Mater. Sci. Eng. R {\bf 58
}, 162 (2007).
\bibitem{Song2014} C. Song and Y. Zheng, J. Colloid Interface Sci. {\bf 427}, 2 (2014).
\bibitem{Guzzardi2007} L. Guzzardi and R. Rosso, J. Phys. A {\bf 40}, 19 (2007).
\bibitem{Hienola2007} A. I. Hienola, P. M. Winkler, P. E. Wagner, H. Venkam\"aki, A. Lauri, I. Napari and M. Kulmala, J. Chem. Phys. {\bf 126}, 094705 (2007).
\bibitem{Cooper2007} S. J. Cooper, C. E. Nichloson, and J. A. Liu, J. Chem. Phys. {\bf 129}, 124715 (2008).
\bibitem{Tao2011} S. Tao, W. Jiadao, and C. Darong, J. Colloid Interface Sci. {\bf 358}, 284 (2011).
\bibitem{Extrand2012} C. W. Extrand and S. I. Moon, Langmuir {\bf 28}, 7775 (2012).
\bibitem{Iwamatsu2015b} M. Iwamatsu, J. Chem. Phys. {\bf 143}, 014701 (2015).
\bibitem{Dobbs1999} H. Dobbs, Physica A {\bf 217}, 36 (1999). 
\bibitem{Brinkmann2005} M. Brinkmann, J. Kierfeld, and R. Lipowsky, J. Phys.: Condense Matter {\bf 17}, 2349 (2005).  
\bibitem{Guzzardi2006} L. Guzzardi, R. Rosso, and E. G. Virga, Phys. Rev. E {\bf 73}, 021602 (2006). 
\bibitem{Mechkov2007} S. Mechkov, G. Oshanin, M. Rauscher, M. Brinkmann, A. M. Cazabat, and S. Dietrich, Europhys. Lett. {\bf 80}, 66002 (2007). 
\bibitem{Indekeu1992}  J. O. Indekeu, Physica A {\bf 183}, 439 (1992). 
\bibitem{Kelton2010} K. F. Kelton and A. L. Greer, {\it Nucleation in Condensed Matter, Applications in Materials and Biology}, Pergamon, Oxford, 2010, Chapter 6.
\bibitem{Fletcher1958} N. H. Fletcher, J. Chem. Phys. {\bf 29}, 572 (1958).
\bibitem{Qian2009} M. Qian and J. Ma, J. Chem. Phys. {\bf 130}, 214709 (2009).
\bibitem{Young1805} T. Young, Phil. Trans. R. Soc. Lond. {\bf 95}, 65 (1805).
\bibitem{Hamaker1937} H. C. Hamaker, Physica {\bf 4}, 1058 (1937).
\bibitem{Kuni1996} F. M. Kuni, A. K. Shchekin, A. I. Rusanov and B. Widom, Adv. Colloid Interface Sci. {\bf 65}, 71 (1996).
\bibitem{Bieker1998} T. Bieker, and S. Dietrich, Physica A {\bf 252}, 85 (1998).
\bibitem{Bykov2006} T. V. Bykov and X. C. Zeng, J. Chem. Phys. {\bf 125}, 144515 (2006).

\end{thebibliography}

\end{document}